# Computational Model for Predicting Particle Fracture During Electrode Calendering

Jiahui Xu[a,b], Brayan Paredes-Goyes[a,b], Zeliang Su[a,b], Mario Scheel[d], Timm Weitkamp[d], Arnaud Demortière [a,b,c], Alejandro A. Franco [a,b,c,e,*]

a. Laboratoire de Réactivité et Chimie des Solides (LRCS), UMR CNRS 7314, Université de Picardie Jules Verne, Hub de l'Energie, 15 rue Baudelocque, 80039 Amiens Cedex, France
b. Réseau sur le Stockage Electrochimique de l'Energie (RS2E), FR CNRS 3459, Hub de l'Energie, 15 rue Baudelocque, 80039 Amiens Cedex, France
c. ALISTORE-European Research Institute, FR CNRS 3104, Hub de l'Energie, 15 rue Baudelocque, 80039 Amiens Cedex, France
d. Synchrotron SOLEIL, 91190 Saint-Aubin, France
e. Institut Universitaire de France, 103 Boulevard Saint Michel, 75005 Paris, France

*Corresponding author: alejandro.franco@u-picardie.fr

## Abstract

In the context of calling for low carbon emissions, lithium-ion batteries (LIBs) have been widely concerned as a power source for electric vehicles, so the fundamental science behind their manufacturing has attracted much attention in recent years. Calendering is an important step of the LIB electrode manufacturing process, and the changes it brings to the electrode microstructure and mechanical properties are worth studying. In this work, we reported the observed cracking of active material (AM) particles due to calendering pressure under ex situ nano-X-ray tomography experiments. We developed a 3D-resolved discrete element method (DEM) model with bonded connections to physically mimic the calendering process using real AM particle shapes derived from the tomography experiments. The DEM model can well predict the change of the morphology of the dry electrode under pressure, and the changes of the applied pressure and porosity are consistent with the experimental values. At the same time, the model is able to simulate the secondary AM particles cracking by the fracture of the bond under force. Our model is the first of its kind being able to predict the fracture of the secondary particles along the calendering process. This work provides a tool for guidance in the manufacturing of optimized LIB electrodes.

## Graphical abstract

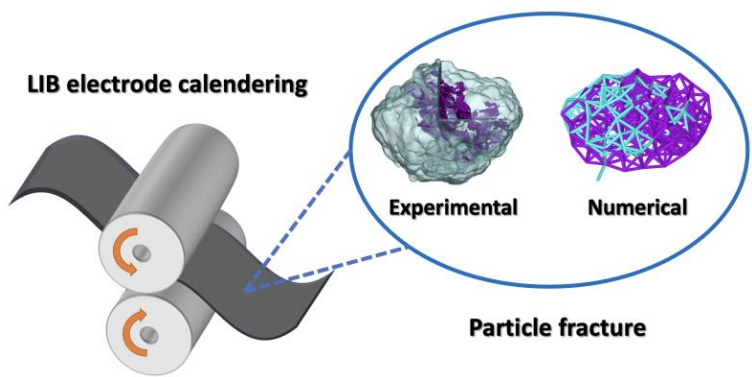

# Introduction

Three decades after the first commercial rechargeable lithium-ion battery (LIB) was introduced, LIBs have become an omnipresent technology in our lives. In the last fifteen years, electric transportation has grown in popularity in response to environmental constraints [1]. However, for private transportation, how to improve the battery energy and power densities while reducing costs is still a problem worthy of research [2]. There is no doubt that material research plays a vital role in finding solutions to this problem, as the energy and power densities, the cycle life, and the safety of LIBs are closely related to the performance of the internal materials. Meanwhile, the manufacturing process is crucial as it impacts the battery quality and cost. The electrode is the most important part in a battery cell since the electrochemical reactions take place here. Its manufacturing process is usually as follows: dry and wet mixing to prepare a slurry, followed by its coating, drying, calendering, dimension control, and final drying [3]. During the calendering step, as the electrode thickness decreases under compression, the porosity decreases and the electronic conductivity increases, resulting in improved electrode volumetric energy density [5]. There is no doubt that the manufacturing parameters of calendering affect the electrode properties. It is reported that pressures, temperatures and the speed of calendering have an impact on electrode performance [6,7], which also depends on the active material (AM) particle size distribution [8–10]. The X-ray Computed Tomography (XCT) has confirmed that the irreversible effects of calendering on electrode morphological change will lead to a significant influence on electrochemical performance [10,11]. Likewise, the deformation and rotation of the AM were observed. Sim *et al.*[12] reported the cracking and fusion in high-nickel cathode materials during calendering by using scanning electron microscope, which alleviated electrochemical capacity decay. AM particle cracking can lead to a deterioration of electrode connectivity and thereby impact electrochemical performance [13], however most of researches have focused on the cracking of AM during electrochemical charge and discharge.

The XCT technique, a state-of-the-art non-destructive method, is increasingly becoming a prominent tool in battery research. By utilizing this technique, researchers can obtain three-dimensional microstructures of battery electrodes, providing valuable insights into their intricate architectures, secondary particle morphologies, and the layout of conductive percolated networks [14–16]. Indeed, the microstructure datasets can be used to investigate and quantitatively study electrode properties such as porosity, tortuosity factor, specific reaction area, anisotropy, and homogeneity. Such work has been used for LIB on materials like graphite [11,14,15], $LiNi_xMn_yCo_{1-x-y}O_2$ (NMC)[11,19,20], $LiFePO_4$ (LFP)[21] and $LiCoO_2$ (LCO) [11,20] both for lab-prepared and commercial LIB. It also enables the tracking of the electrode size, shape, and network morphology changes like AM cracking and irreversible deformation due to mechanical factors during and after the manufacturing process [10]. Additionally, a variety of algorithms based on machine learning have been developed to realize the automatic segmentation of each material phase (AM, carbon-binder and pores) in the grayscale image dataset [22,23].

Computational models are popular in this digital age and offer a plethora of interesting capabilities to gain insights about the electrode microstructure. To link with industrial production, digital models for different steps of the LIB manufacturing process have been developed [24–26]. Several studies have been done by computational simulation to investigate the LIB electrode architecture changes during the manufacturing process. Sangros Gimenez et al. [27] simulated the electrode behavior in the calendering process by a discrete element method (DEM) approach by describing explicitly only the active material particles.

Srivastava et al. [28] predicted the electrode properties by controlling binder adhesion during manufacturing process by using a combination of discrete element and colloidal dynamics methods. Nikpour et al. [29,30] developed the multi-phase smoothed particle model to study the electrode heterogeneity and the electrode properties. Ge et al. [31,32] performed calendering DEM calculations using X-ray tomography data of electrode as initial microstructures for the simulations. They represented the binder effect by employing a bond model between AM particles. Lippke et al. [33] used DEM to identify the impact of the electrode preheating on the calendering process. Wang et al. [34] used a DEM with an additional bond model to describe the mesostructure evolution under stress and discover the mechanical integrity of the active layer is influenced by the binder content and the active particle size distribution. Asylbekov et al. [35] used a microscale Continuum Fluid Dynamics (CFD)-DEM to investigate the breaking behavior of carbon black aggregates due to the shear stress during the mixing process. Lundkvist et al. [36] developed a method to simulate the electrode creation and calendering with DEM, which is able to capture the unloaded stiffness behavior induced by the viscoelastic binder. All these works (except references 29, 30) do not account for an explicit representation of the carbon-binder domain phase. In our ARTISTIC project[37], funded by the European Research Council, we developed a series of pioneering sequentially coupled 3D-resolved physical-based models to represent the different steps of the LIB electrode manufacturing process by using coarse-grained molecular dynamics and DEM. This digital twin of the electrode manufacturing process, allows to predict, as a function of the manufacturing process parameters, the spatial location of both active and inactive phases in a 3D-resolved fashion [38–43].

Earlier, we reported the introduction of real particle shapes, obtained by XCT, into our ARTISTIC manufacturing process model based on coarse-grained molecular dynamics (CGMD) and DEM [44]. In this model, the secondary AM particles, consisting of spherical primary particles, represent the NMC111 material in the electrode manufacturing processes. Spherical particles are used to mimic the carbon binder domain (CBD) consisting of carbon, binder and nanopores. In this work, we performed an ex situ nanoscale XCT experiment on the electrode before and after calendering process and assessed the fracture inside the secondary particles. Therefore, we used and optimized the existing model to further investigate the mechanical behavior of the electrode and probe the effect of calendering pressure on the particle cracking.

# Experimental Analysis

## Methods

### Sample preparation

$LiNi_{0.33}Co_{0.33}Mn_{0.33}O_2$ (NMC111) (Umicore), carbon black (C-NERGY™ super C65) and Polyvinylidene fluoride (PVdF) (Solef™ 5130/1001, Solvay) are premixed in a weight ratio of 96:2:2 for overnight. Afterwards, NMP is added until reaching the desired solid content. The mixture is performed in a Dispermat CV3-PLUS high-shear mixer for 2 h at 25 °C. Then, the slurry is coated over a 22.5 μm thick Aluminum current collector by using a comma-coater prototype-grade machine (PDL250, People & Technology, Korea) and passes through two ovens with temperatures of 85 °C and 95 °C. After drying, the electrodes are calendered in a prototype-grade lap press calender (BPN250, People & Technology, Korea) under various roll gaps and at constant line speed of 0.54 m/min at 60 °C. The calendering pressures were transformed from a calculation by using the roll gaps and the thickness of the electrode, which was calibrated by measuring the applied force of the corresponding roll gap through the utilization of a force sensor film (ELF measuring system equipped with FlexiForce sensors, Tekscan)[6].

Small electrode pieces were cut from the center of the electrode and glued horizontally by epoxy onto the metal needles. The tops of the electrode were cut by a laser-equipped microdissector into the size that corresponds to the field of view of the experiment.

## Nano X-ray Computed Tomography

The nano X-ray tomographic analysis of the calendered electrode under different compression pressures were performed at the beamline Anatomix of Synchrotron SOLEIL, France[45] where transmission X-ray microscopy (TXM) instrument was used in mode absorption[46]. The scans were conducted with a central photon energy of 8.327 keV using a TXM configuration. 1000 projections are acquired over 360° with an exposure time of 2 s per 2D image. Volumes were reconstructed with PyHST2[47] (ESRF, Grenoble, France). A paganin filter was applied during the data reconstrction [48]. The 3D volumes with a pixel size of 41 nm were achieved after reconstruction.

To reduce or eliminate the noise and blurriness caused by image artifacts, the images were preprocessed prior to analysis. A non-local means filter was applied, followed by an unsharp mask to enhance the edges of all tomographic data. Then the machine learning segmentation plugin Trainable Weka [49] based on random Forest algorithm in FIJI software suit had been used to segment the 3D data into different phases. The methodology involved extracting features from images using various filters and a random forest algorithm for classification.

## Results and discussion

The electrode of a thickness of 114 µm was calendered between roll gaps of 21 µm and 6 µm. The respective pressures, the mass and thickness measured, and porosity calculated are shown in Table 1.

Table 1: The experimental information of the electrodes.

| Roll gap (µm) | Corresponding pressure (MPa) | Mass (mg) | Thickness (µm) | Porosity (%) |
|---|---|---|---|---|
| - | 0 | 22.37 ± 0.29 | 99.03 ± 0.87 | 47.97 ± 0.87 |
| 21 | 80 | 22.97 ± 0.11 | 79.67 ± 0.06 | 27.75 ± 0.23 |
| 6 | 150 | 22.30 ± 0.68 | 75.4 ± 0.79 | 25.17 ± 1.2 |

The 2D slice images of these electrodes obtained from the nano-XCT are shown in Figure 1 (uncalendered and calendered under 150 MPa) and Figure S1 (calendered under 80 MPa). It can be clearly seen that the AM particles before calendering have already pores caused by material synthesis, but the particle structure is intact without cracks. As the pressure on the electrodes increased, the particles remained relatively intact at 80 MPa. At a pressure of 150 MPa, the particles showed obvious cracks.

The 3D images with gray values of the electrode before calendering and calendered under 150 MPa were segmented into 3 phases: AM, macro pores and the nano-cracks. The triphasic image stacks were imported into the commercial software package Avizo V9.4 (Thermo Fisher Scientific) [50] for visualization. The 3D visualization is presented in Figure 1 and the voxel size in the 3D particles volumes is $41 \times 41 \times 41$ nm$^3$. Among the 3D volumes, only the AM phase and nano-cracks are visualized, and the macro pore phase is presented as void in the volumes. As shown in Figure 1, it is obvious that the calendering process makes a great difference on the electrode density. Several individual particles were selected to demonstrate the intrinsic pore inside the AM particles and the nano-cracking due to the high pressure during the calendering process. The 3D visualization clearly shows the presence of cracks within the secondary particles after compression and the cracks are not evenly distributed or exist in every secondary particle. However, due to the small size of our electrodes volume, the cracks could not be accurately quantified. Another micro-scale ex situ XCT experiment was done to study the change of porosity and tortuosity factor of the electrode after calendering, the results are shown in Figure S2. Due to experimental limitation, it is difficult to obtain *in operando* the mechanical behavior of particles under each pressure. We therefore developed a new physical model to predict the fracture of secondary particles during the calendering process.

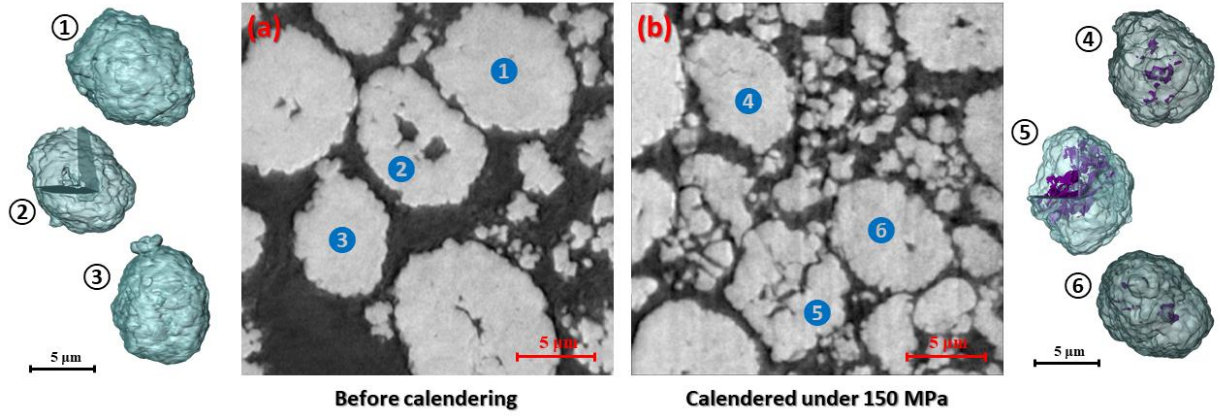

Figure 1: Nano XCT. 2D and 3D visualization of the electrode (a) before calendering; (b) calendered under 150 MPa. The individual particles are selected and presented in 3D. The surfaces of the particles and the nano cracking are indicated in turquoise and violet, respectively. Particles 2 and 5 are sectioned to reveal internal structures. The scale bars in the figure are equal to 5 μm.

## Numerical Analysis

### Methods

The calendering process is modeled at the particle scale using the Discrete Element Method [51]. In addition to AM, we explicitly consider the carbon binder domain (CBD), which allows the obtention of more reliable mesostructures[41]. The CBD domain is represented by spherical particles with a diameter of 1.16 µm with a nanoporosity of 27%. As in our previous work [44], the secondary AM particles present both realistic shape and particles size distribution (PSD) obtained by X-ray microcomputed tomography. These secondary particles are composed of spherical AM primary particles. The calender rolls and current collector are represented by a top and a bottom plane respectively, more information can be found in our earlier work [41]. The simulation box has a periodic boundary condition in the direction parallel to the plane and the current collector.

The new position of each particle is determined by the contact and external body forces, in our case the latter consisting only of gravity. Every pair of particles in contact experience an elastic Hertzian force composed of normal and tangential components:

$$\boldsymbol{F}_E^n = k_n \delta \, \boldsymbol{n} - \gamma_n \boldsymbol{v}_r^n \tag{1}$$

$$\boldsymbol{F}_E^t = -min(\mu_t |\boldsymbol{F}_E^n|, |-k_t \boldsymbol{S_t} - \gamma_t \boldsymbol{v}_r^t|) \, \boldsymbol{t} \tag{2}$$

where $k_n$, $k_t$ are the normal and tangential elastic constants, while $\gamma_n$, $\gamma_t$ are the normal and tangential viscoelastic damping constants. These constants are estimated from the following particle properties: Young's modulus $E_p$, Poisson's ratio $Po_p$ and restitution coefficient $e_p$. $\delta$ is the overlap distance, $\boldsymbol{v}_r^n$, $\boldsymbol{v}_r^t$ the normal and tangential relative velocities and $\boldsymbol{S_t}$ the accumulated tangential displacement. $\boldsymbol{n}$, $\boldsymbol{t}$ are the normal and tangential unit vectors.

The secondary AM particles and the CBD particles are also subjected to cohesion forces, here represented by a Simplified JKR model (SJKR). This force is an attractive normal force given by:

$$\boldsymbol{F_C} = -c_{CED}\, A_{COH}\, \boldsymbol{n} \tag{3}$$

where $c_{CED}$ is the cohesion energy density and $A_{COH}$ is the contact area (sphere-sphere). On the other hand, the existing sintering force between primary particles belonging to the same secondary particle is represented by the bonded-particle model (BPM)[52]. The normal and tangential incremental forces of a bond are calculated at each timestep following [53]:

$$\Delta F_B^n = \frac{E_b A_b}{l_b} v_r^n \Delta t \tag{4}$$

$$\Delta F_B^t = \frac{(E_b/2(1+Po_b))A_b}{l_b} v_r^t \Delta t \tag{5}$$

where $A_b$, $l_b$ are the cross-section area and the length of the assumed cylindrical bond. $E_b$, $Po_b$ are the Young's modulus and the Poisson's ratio of the bond. In order to decrease the free parameters, these two bond properties are set to the same value of the particle parameters. The latter ($E_p$, $Po_p$) are straightforward calibrated by fitting the maximum pressure during calendaring.

Similar expressions can be found in [53] to calculate the axial $\Delta M_B^n$ and shear $\Delta M_B^t$ incremental moments of the bonds. In this way, the beam theory allows to calculate the maximum normal and shear stresses at the bond following [52]:

$$\sigma_B^n = \frac{-|F_B^n|}{A_b} + \frac{|M_B^t| R_{min}}{I} \tag{6}$$

$$\sigma_B^t = \frac{|F_B^t|}{A_b} + \frac{|M_B^n| R_{min}}{J} \tag{7}$$

where $R_{min}$ is the lower of the radii of particles $i$ and $j$, while $I, J$ are the moment of inertia and the polar moment of inertia of the cross section of the bond.

Following the failure criteria proposed by the BPM, a bond breaks if the maximum normal stress is higher than the bond normal strength ($\sigma_B^n \geq \sigma_B^{cn}$) or if the maximum tangential stress is higher than the bond shear strength ($\sigma_B^t \geq \sigma_B^{ct}$). Other failure criteria can be found in [54]. For normal and shear breakage to be possible, $\sigma_B^{cn} = \sigma_B^{ct}$[52]. This strength value at the bond level should be found by calibration. Here, as a first DEM model for the fracture of NMC particles, a simply parametric study is performed to elucidate the influence of this value.

The initial electrode structure for calendering was derived from the simulation result of drying in our previous work[44] to maintain continuity in our manufacturing simulations. The real shape of the secondary particles in the simulation were extracted from the micro-scale XCT. The methods used for the image digitization and the initial structure generation is described in detail in the previous paper, as well as the slurry and drying simulation detail. The computational workflow is shown in Figure 2. Bonds are created between AM primary particles (of the same secondary particle) that are in contact in the initial microstructure.

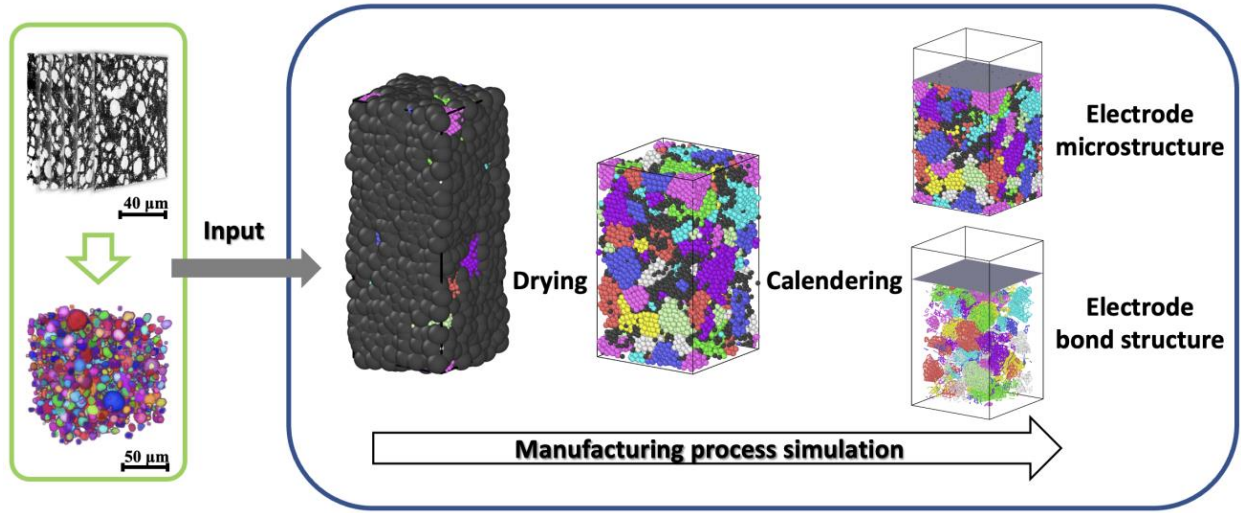

Figure 2: Computational workflow. In the simulation box, the black refers to the CBD particles and the rest colors indicate the different secondary AM particles. In the scheme for calendering process, the bonds in the secondary particle were presented with the same color as the particle.

The slurry and drying simulation are performed by using the open-source molecular dynamics simulator software LAMMPS. The real shape of the NMC material was obtained through the micro-XCT experiment. After reconstruction, image processed, particles labeling, the individual particles were stored in a matrix and imported into the initial simulation box by using the powder particle size distribution. The detail of these steps and the slurry and drying simulation can be found in our previous paper[44]. The simulations are performed in the open-source DEM software LIGGGHTS[55], using a modified version[56] where the BPM is implemented. The speed of the top plane is 0.01 m/s, close to the experimental line speed 0.009 m/s. We verified negligible influence of this speed on the obtained pressures. A timestep of 0.01 ns (around 1.5% of Rayleigh time) was verified to give numerical stability. The simulation duration is from 1 to 4 days depending on the degree of compression. These were run in 28 cores Intel(R) Xeon(R) CPU E5-2680 v4 @ 2.40GHz in 1 node (128 GB of RAM) of the MatriCs platform (Université de Picardie-Jules Verne, France).

## Results and discussion

The parameterization of the model is performed by pressure-porosity curves resulting from experimental results as shown in Table 1. The maximum pressure during calendering is recorded in the simulation, while the porosity is calculated from the electrodes after relaxation. The calendering process is mimicked by two steps: compression and relaxation. The upper plane first moves downward with the given velocity to a certain displacement to reach the maximal compression and then moves back to the original location with the same velocity so as to achieve the relaxation. The relaxation step is a representation of the fast elastic recovery of the electrode as reported in the literature[57]. Figure 3 shows one electrode before and during the simulated calendering process under several pressures and the corresponding bond structures, which correspond to uncalendered, calendered under experienced pressures of 9.2 MPa, 20.1 MPa, 50.9 MPa and 146.8 MPa. During the calendering process, the nanoporosity of CBD decreased from 50% at the beginning to 27% under pressure, which is similar to the porosity reported in literature[58]. Our model introduced bonds to refer to the connection between the primary particles within the secondary particles. Although the real primary particle crystal shape is not spherical and the size is smaller than the primary particle size within the model, here, we perform a coarse-grained simulation with primary spherical particles of 1.59 μm diameter to simulate the AM mechanical behavior against the strain. An individual particle is singled out to demonstrate its change during calendering, as shown in Figure 5. At the given failure criteria, the bonds are broken to represent the disconnection between primary particles. According to the evolution of the bond structure in Figure 3, it can be seen that the bonds break gradually with the increase of the applied pressure. However, the location of the bond breakage is not uniformly distributed in the electrode. This location is related to the distribution of the particles in the electrode and the force applied. Also, the number of bond breaks is not linearly related to the reduction of the thickness due to the uneven force inside the electrode, as shown in the Figure 4b. To the best of our knowledge, this is the first time that DEM simulations have been used to predict the effect of the calendering process on secondary particle fracture.

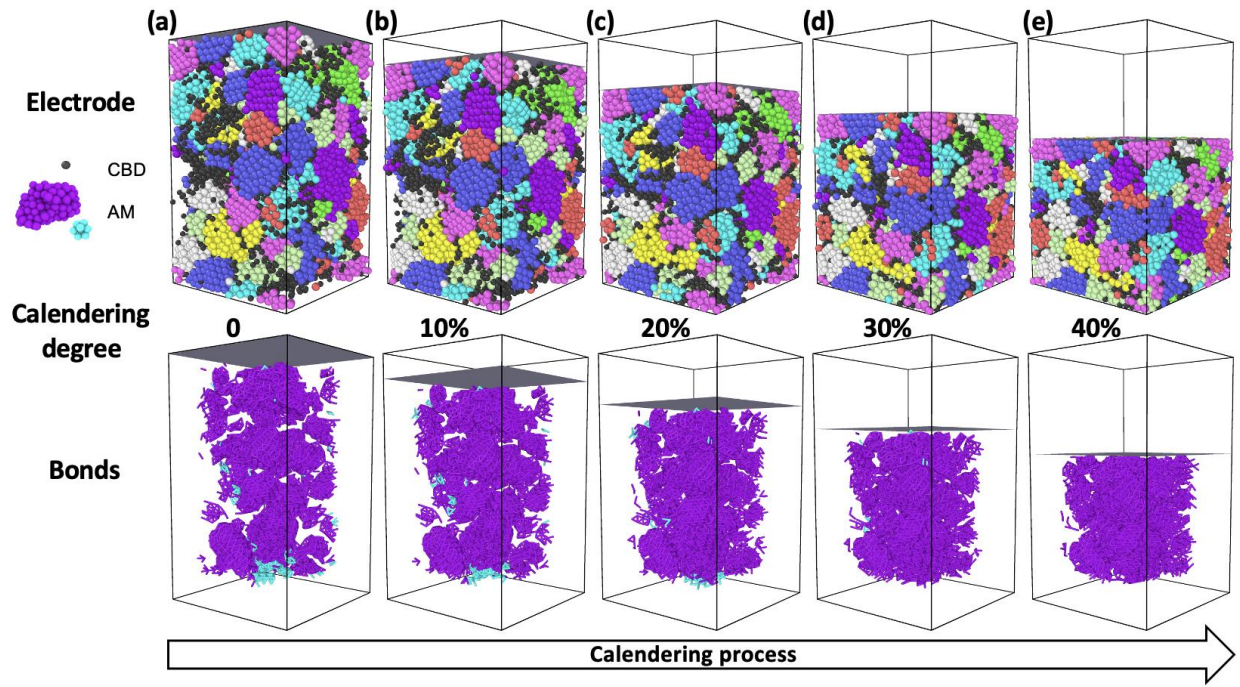

Figure 3: The electrodes and bonds evolution during the calendering process of a maximum pressure of 150 MPa. The presented structure is the structure under the pressure of (a)uncalendered, (b)calendered under 9.2 MPa, (c)20.1 MPa, (d)50.9 MPa and (e)146.8 MPa during the compression process. In the electrode structure, black indicates the CBD and the rest of the colors represent the different AM secondary particles. In the bond structure, blue indicates the bonds that will break during this compression, while purple indicates bonds that remain intact throughout.

In this work we use the experimental porosity vs. pressure profile to validate our model. As shown in Figure 4a, the comparison of the porosity between the model results and the experimental ones at different calendering pressures were obtained. The porosity of the electrode from the simulation is calculated from the volume after relaxation process. Overall, our simulation results have a good agreement with the experimental results, demonstrating the connection between mechanical behavior and microstructure evolution of the electrodes, solidifying our fracture prediction. Figure 4b demonstrates the relationship between the displacement and the applied pressure. The pressure was calculated by the contact force and the contact surface area between the particles and the upper plane, the latter of which is 910 µm$^2$. The blue curve indicates the compression process, and the red is the relaxation process. The pressure-displacement curves are consistent with the results of previously reported micro-indentation experiments [41]. The green curve represents the proportion of broken bonds to the initial total number of bonds at this displacement. From the figure, it is concluded that approximately 6.8%, 17.8% and 20.3% of the bonds are broken when the pressure reached 20 MPa, 90 MPa and 146 MPa, respectively. Combined with our Nano-XCT images observed at 150 MPa (Figure 1b) and the individual particle study (Figure 5d), we can speculate that in the model, secondary particle cracking occurs when the number of broken bonds is approximately 8%-12%, i.e., when the applied pressure is approximately 50 MPa-140 MPa.

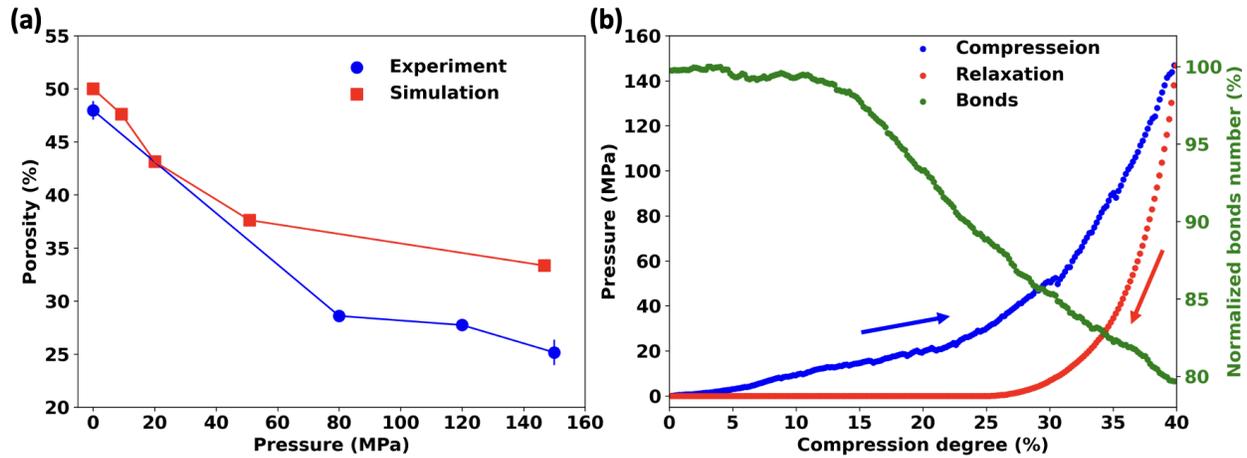

Figure 4: Mechanical and porosity validation. (a)Comparison of porosity between model results and experimental value at different calendering pressures. (b)Pressure and proportion of bond breakage vs. upper plane displacement (reduced thickness).

From our experimental observations not all secondary particles have fractures under this pressure, and here we have selected one of the most likely fractured secondary particles to study the evolution of its shape and internal structure. Since the nano-fracture we observe is smaller than the scale of our model, we are not able to observe the fracture of the secondary particle in the model by visualizing the output of the model. Figure 5 shows the evolution of an AM particle with an equivalent diameter of 11.96 μm during the calendering process under 146 MPa. It can be seen that there is a slight deformation of the particle due to the force, which is accompanied by the bond breakage. According to our hypothesis that 13% of fracture leads to nano-cracking, the particle is accompanied by nano-cracking during the deformation. Due to the modeling scale limitation, we cannot visualize the fracture and the integration of the binder and carbon additive into it, which, according to the literature [12], is a possible occurrence and contributes to the electrode's resistance to the capacity decay.

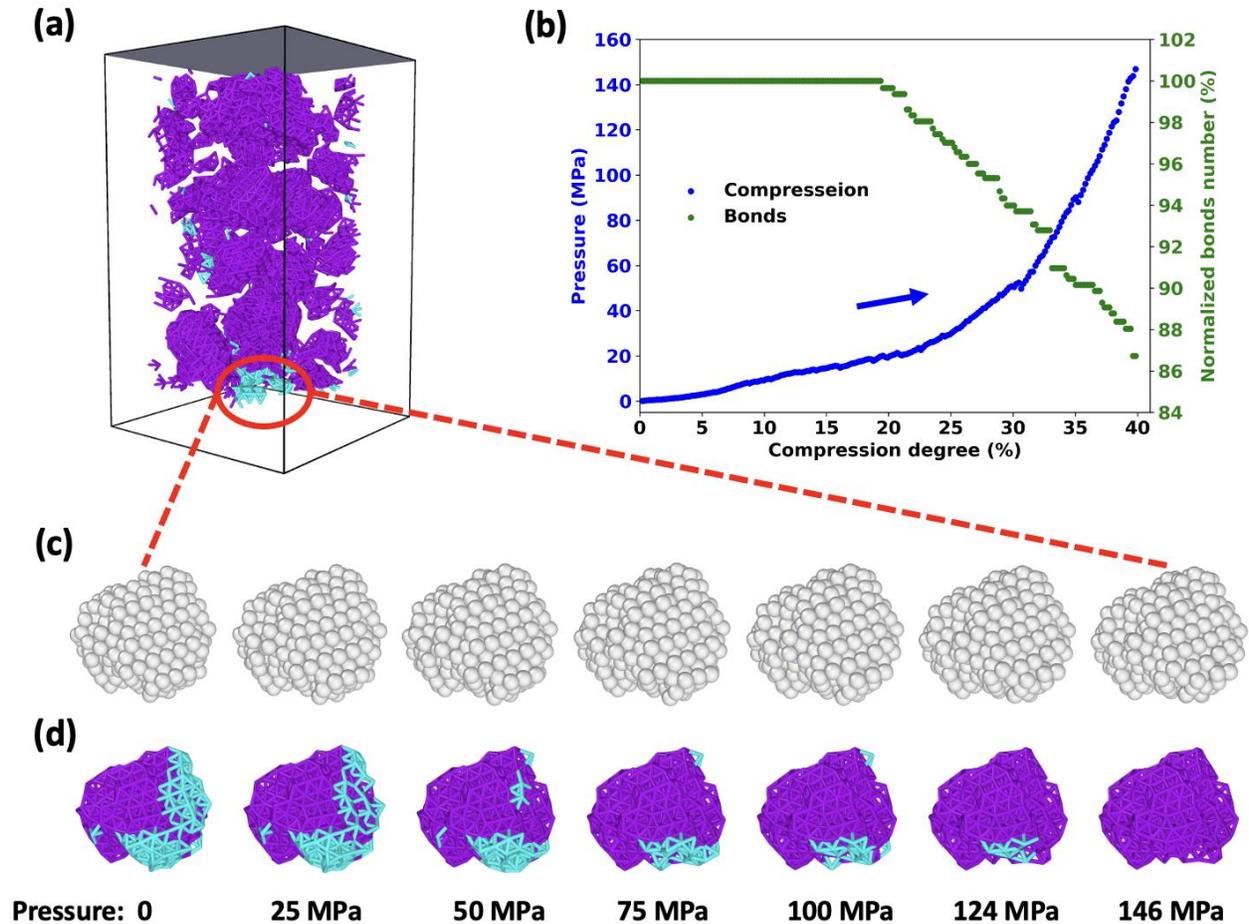

Figure 5: Study of selected individual particles with large ratio of bond breaks and possible cracks. (a)The location of the selected particle in the volume. (b)The pressure applied on the simulated electrode and number of bonds break of this individual particle. (c) Particle shape evolution during the calendering. (d)The change of the bonds within the selected particles during the calendering. The blue indicates the bonds that will break during this compression, while purple indicates bonds that remain intact throughout.

# Conclusions

In this work, we have first precisely assessed the fracture of NMC111 secondary particles due to pressure during the calendering process of the LIB electrodes through ex situ nano-XCT experiments, and the cracks are 3D visualized. From this, we developed a 3D LIB electrode calendering model using DEM based on the real AM particle shape obtained by micro XCT to predict the mechanical behavior of the particles under pressure during compression. The initial structure was derived from the results of our previous work using a coarse-grained molecular dynamics-based physical model for simulating slurry and solvent evaporation. In this new work, bonds between primary particles are introduced into the structure to simulate the connection of the primary particles. This work reports the calendering and the relaxation of a single formulation under different pressure conditions in the range of 0-150 MPa, and the obtained electrode porosity for the simulation are in good agreement with the experimental results. It is found that the breakage of the bonds does not depend linearly on the applied force. With the increase of the pressure and the decrease of the electrode porosity, the source of the microstructural change in the electrode changes from the movement of particles to the deformation and fracture of particles. This new approach allows us to focus on the generation of internal cracks in individual particles, and by calibrating the criteria for the number of breakages of the bonds for secondary particle breakage using XCT experiments and computational simulation, the model makes it possible to predict the generation of cracks in secondary particles. Our work continues the goal of the ARTISTIC project to more realistically and detailly simulate the process of the LIB manufacturing. The microstructures obtained from the model will be used to simulate the electrochemical performance in the future to capture the subtle effects of particle deformation and fracture on the electrochemical performance. Our model can also be applied to other materials and chemistries, which we believe will be instructive for the entire battery community and industry.

## Authorship contribution statement

**Jiahui Xu:** Conceptualization, Methodology, Software, Experiment, Visualization, Writing - original draft.

**Brayan Paredes-Goyes:** Conceptualization, Methodology, Software, Writing – original draft.

**Zeliang Su**: Experiment, Visualization.

**Mario Scheel**: Experiment, Visualization, Writing - review & editing.

**Timm Weitkamp**: Synchrotron access.

**Arnaud Demortière:** Co-supervision, Writing-review & editing.

**Alejandro A. Franco:** Conceptualization, Methodology, Supervision, Funding acquisition, Project administration, Resources, Writing - review & editing.

## Declaration of competing interest

The authors declare that they have no known competing financial interests or personal relationships that could have appeared to influence the work reported in this paper.

## Acknowledgments

A.A.F. and J.X. acknowledge the European Union's Horizon 2020 research and innovation program for the funding support through the European Research Council (grant agreement 772873, "ARTISTIC" project). A.A.F. and B.M.P. acknowledge the European Research Council for the funding support through the ERC Proof-of-Concept grant No 101069244 (SMARTISTIC). A.A.F. acknowledges Institut Universitaire de France for the support. The authors acknowledge the MatriCS HPC platform from Université de Picardie-Jules Verne for the support and for hosting and managing the ARTISTIC dedicated nodes used for the calculations reported in this manuscript.